\documentclass[fleqn,usenatbib]{mnras}

\usepackage{newtxtext,newtxmath}

\usepackage[T1]{fontenc}

\DeclareRobustCommand{\VAN}[3]{#2}
\let\VANthebibliography\thebibliography
\def\thebibliography{\DeclareRobustCommand{\VAN}[3]{##3}\VANthebibliography}

\usepackage{graphicx}	
\usepackage{amsmath}	

\title[Polarization of Cygnus X--1]{X-ray Polarization changes with the State Transition in Cygnus X--1}

\author[A. Jana et al.]{
Arghajit Jana,$^{1, 2}$\thanks{E-mail: arghajit.jana@mail.udp.cl; argha0004@gmail.com}
Hsiang-Kuang Chang$^{1}$\thanks{E-mail: hkchang@mx.nthu.edu.tw}
\\
$^{1}$Instiute of Astronomy, National Tsing Hua University, Hsinchu 300044, Taiwan \\
$^{2}$Instituto de Estudios Astrof\'isicos, Facultad de Ingenier\'ia y Ciencias, Universidad Diego Portales, Av. Ej\'ercito Libertador 441, Santiago, Chile
}

\date{Accepted XXX. Received YYY; in original form ZZZ}

\pubyear{2023}

\begin{document}
\label{firstpage}
\pagerange{\pageref{firstpage}--\pageref{lastpage}}
\maketitle

\begin{abstract}
We studied the polarization properties of Cygnus X--1 in both hard and soft spectral states with Imaging X-ray Polarimetry Explorer (IXPE) observations. The polarization degree is $\sim 4\%$, and $\sim 2.6-2.8\%$ in the hard and soft states, respectively. The polarization angle is observed along the jet axis and remains the same in both states. Energy-dependent analysis revealed an increasing polarization with the energy. The observed polarization rules out the lamp-post corona and prefers a conical corona if the coronal geometry remains the same in the hard and soft states. The disk is not detected in the hard state but is found in the soft state in the IXPE spectra. The spectral analysis revealed that the disk emission contributes $\sim 20\%$ of the total emission in the soft state. The scattering of the seed photons inside the corona is likely to reduce the polarization. This could explain the different polarization in different spectral states. Additionally, if the disk polarization is perpendicular to that of the corona, it could also reduce the polarization in the soft state.
\end{abstract}




\section{Introduction} \label{sec:intro}

Cygnus X-1 is the first black hole X-ray binary (BHXB) discovered in 1971 \citep{Oda1971,Tananbaum1972}. Cygnus X--1 consists of a black hole of mass $21.2\pm2.2~M_{\odot}$ \citep{Miller-Jones2021}, and a O-type super-giant star of mass $40.6^{+7.7}_{-7.1} M_{\odot}$ \citep{Miller-Jones2021}. The black hole in Cygnus X--1 is found to be a highly spinning BH, with spin parameter $a^* >0.99$ \citep[e.g.,][]{Gou2014,Tomsick2014,Zhao2021}. The binary system is located at a distance of $2.2\pm0.2$ kpc \citep{Miller-Jones2021}, with an orbital period of 5.6 days. The orbital inclination angle is reported to be $27^\circ{}.1 \pm 0.8$ from the optical observations \citep{Orosz2011}.

Cygnus X--1 mostly remains in the non-thermal emission-dominated hard spectral state. However, occasionally, it transits to the thermal emission-dominated soft state \citep[e.g.,][]{Tomsick2014,Basak2017,Ghosh2019}. The thermal emission is believed to originate in a geometrically thin, optically thick disk \citep{SS73}. In contrast, the non-thermal emission is thought to originate from a hot electron cloud, known as corona \citep[e.g.,][]{ST80,HM93}. The thermal seed photons from the accretion disk produce a multi-color blackbody component. A fraction of thermal photon is intercepted in the corona and produces non-thermal hard powerlaw emission via Compton up-scattering with the hot electrons \citep[e.g.,][]{T94,CT95,Done2007}. Some authors suggest that the jet also contributes to the high energy emission, especially above 100 keV \citep[e.g.,][]{Rahoui2011,Russell2014}.

Over the years, Cygnus X--1 was studied to understand the accretion geometry. However, the geometry of the corona is still not understood, and highly debated. Several models with different coronal geometries, such as lamp-post, spherical, or slab geometry, have been considered. The spectral study alone cannot infer the geometry of the corona. One must carry out an X-ray polarimetric study to constrain the coronal geometry. The thermal photons are expected to be polarized as they suffer electron scattering in the inner disc region. As the thermal photons are up-scattered in the corona to produce the hard photons, the Comptonized emission would also be polarized \citep[e.g.,][]{Stark1977,Schnittman2009,Schnittman2010}.

The first polarization observation of Cygnus X-1 was carried out with the OSO--8, which gave the polarization degree (PD) as $2.44\pm1.07\%$ and the polarization angle (PA, $\psi$) as $-18^\circ{} \pm 13$ (measure on the plane of the sky, from north to east) at 2.6 keV \citep{Long1980}. INTEGRAL observation reveals a high PD as $65\%$ with the PA as $224^\circ{}$ \citep{Laurent2011,Jourdain2012} in the 0.4--2 MeV energy range. AstroSat/CZTI observation suggested the PD as $\sim 23\%$ and PA as $236^\circ{}$ at $100-380$~keV energy range in the hard-intermediate state \citep{Chattopadhyay2023}. \citet{Chauvin2018} measured an upper limit of the PD as $8.6\%$ (at 90\% confidence) and PA along the jet axis in the $19-181$ keV energy range with PoGO+ observations. In the low energy band, \citet{Krawczynski2022} reported the PD as $4.0\pm0.2 \%$ and PA as $-20^\circ{}.7\pm1.4$ in the $2-8$ energy range with the IXPE observation in the hard state. 

The Imaging X-ray Polarimetry Explorer \citep[IXPE;][]{Weisskopf2022} observed Cygnus X--1 five times between May 2022 and May 2023. Of the five observations, two in 2022 were carried out in the hard state, while three in May 2023 were carried in the soft state. In this paper, we carried out a polarimetric study of Cygnus X--1 using the IXPE observations in the hard and soft states. We also used the simultaneous data obtained from NICER and NuSTAR for the spectro-polarimetric study. The paper is organized as follows. Section~\ref{sec:obs} describes the data reduction process. In Section~\ref{sec:analysis}, we presented the analysis process and our findings. Finally, in Section~\ref{sec:dis}, we discuss our findings and summarize our result.

\section{Observation and Data Extraction} \label{sec:obs}

IXPE consists of three identical detector units (DUs) operating in the $2-8$ keV energy ranges. IXPE observes Cygnus X--1 five times between May 2022 and May 2023. The observation log is tabulated in Table~\ref{tab:log}. We used calibrated and cleaned level2 data for our analysis, which is supplied by the Science Operation Center (SOC) using \textsc{ixpeobssim} software v30.0.0 \citep{Baldini2022}. The source and background products were extracted by considering circular regions with 60\,arcsec, centered at the source coordinates and away from the source, respectively, using \textsc{xpselect} task. Then, the polarization cube were constructed with the \textsc{pcube} algorithm \citep{Kislat2015} with \textsc{xpbin} task. The Stokes I, Q, and U source and background spectra are generated using PHA1, PHA1Q, and PHA1U algorithms. In all the observations, the background accounted for $<0.1\%$ of the source count rate.

The Neutron Star Interior Composition Explorer Mission (NICER) observed Cygnus X--1 simultaneously with the IXPE in May 2022, June 2022. The \textsc{nicerl2} task is used to perform standard calibration and screening to generate cleaned event files. The source and background spectra with responses were generated using the \textsc{nicerl3-spect} task. The spectra are rebinned to have minimum 25 counts per bin using \textsc{grppha} task.

The Nuclear Spectroscopic Telescope Array (NuSTAR) observed Cygnus X--1 simultaneously with IXPE three times. NuSTAR is a hard X-ray focusing telescope with two identical modules, FPMA and FPMB, and operates in the $3-78$~keV energy range \citep{Harrison2013}. The data were reprocessed with the NuSTAR Data Analysis Software (\textsc{nustardas}\footnote{\url{https://heasarc.gsfc.nasa.gov/docs/nustar/analysis/}}, version 1.4.1). Using standard filtering criteria, we generated clean event files with the \textsc{nupipeline} task. The data were calibrated using the latest calibration data files available in the NuSTAR calibration database\footnote{\url{http://heasarc.gsfc.nasa.gov/FTP/caldb/data/nustar/fpm/}}. The source and background products were extracted by considering circular regions with 60\,arcsec, and 90\,arcsec radii, centered at the source coordinates and away from the source, respectively. The spectra were extracted using the \textsc{nuproduct} task and then re-binned to ensure that they had at least 25 counts per bin using the \textsc{grppha} task. 

Throughout the paper, We quoted the errors at 1.6 $\sigma$ level (90\% confidence), unless stated.

\begin{figure}
\centering
\includegraphics[width=8.5cm]{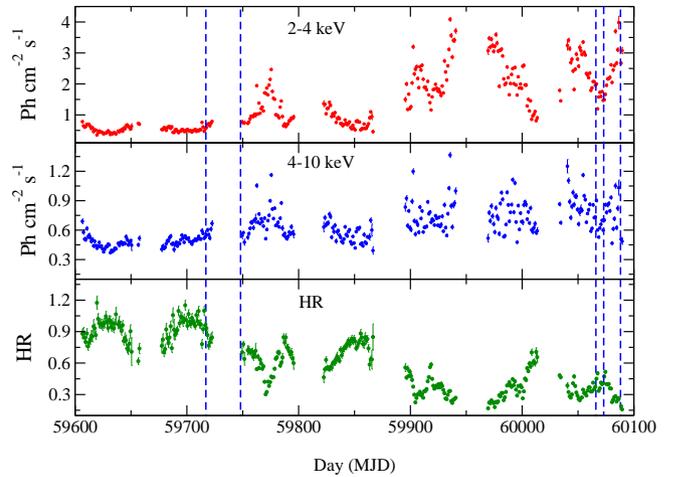}
\caption{Lightcurves of Cygnus X--1 are shown, obtained by MAXI. The variation of count rate in the $2-4$~keV and $4-10$~keV are shown in the top and middle panel, respectively. In the bottom panel, the variation of hardness ratio (HR) is shown. The vertical dashed lines mark the IXPE observations.}
\label{fig:lc}
\end{figure}

\section{Analysis and Results} \label{sec:analysis}
\subsection{Lightcurve \& Spectral State}\label{subsec:lc}
Figure~\ref{fig:lc} shows the lightcurves of Cygnus X--1, obtained by MAXI. The variation of count rate in the $2-4$~keV and $4-10$~keV are shown in the top and middle panels, respectively. The variation of hardness ratio (HR) is shown in the bottom panel. The HR is defined as the ratio of count rate in $4-10$~keV energy range to the count rate in the $2-4$~keV energy range. The vertical dashed lines represent the IXPE observations. During the first two observations (H1 and H2; see Table~\ref{tab:log}), the $2-4$ count rate was low, with a high HR as Cygnus X--1 was in the hard state. In the last three observations, Cygnus X--1 transits to the soft state as the $2-4$~keV count rate increases, with the HR decreases. Here, we note that the spectral states of Cygnus X--1 is not the same as other BHXRBs \citep[e.g.,][]{Gierlinski1999,Walton2016,Yan2021}.

\begin{figure*}
\centering
\includegraphics[width=7cm]{stokes-param.eps}
\includegraphics[width=8cm]{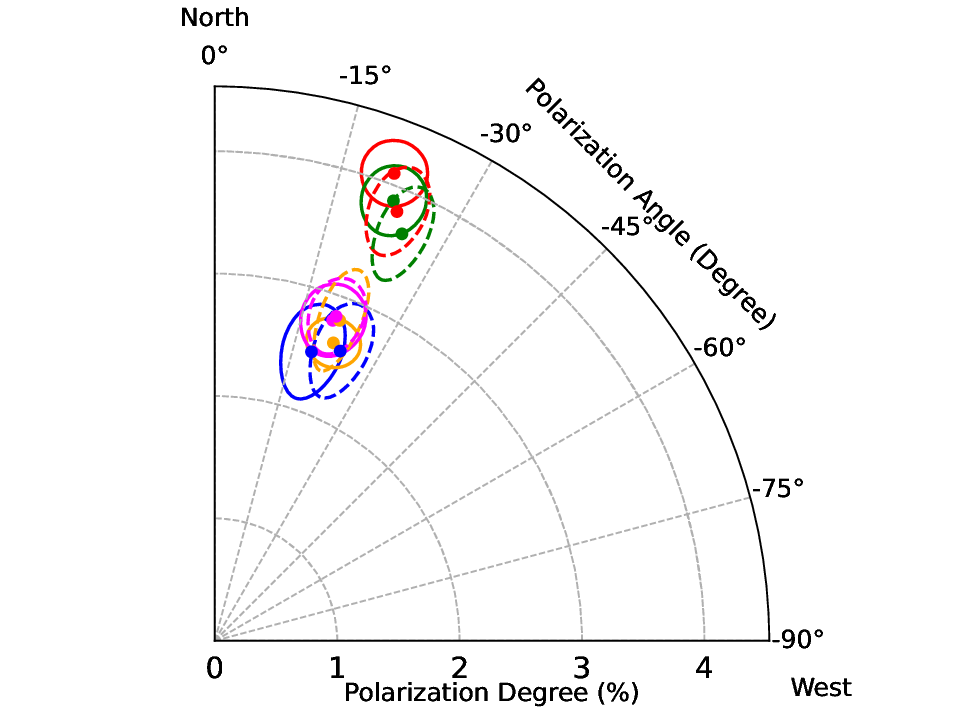}
\caption{Left panel: Stoke parameter Q/I and U/I for all the observations for all three DUs in the $2-8$~keV energy bands. Different color represent the different observations. 
Right panel: Contours of PA and PD, obtained from the \textsc{pcube} algorithm, and spectro-polarimetric analysis in \textsc{xspec} with the \textsc{polconst} model. The eclipses with solid and dashed lines represent the contour from the \textsc{pcube} algorithm, and the spectro-polarimetric analysis with in \textsc{xspec}. The errors are plotted in 3$\sigma$ level. The red, green, blue, orange and magenta points and eclipses represent the data from H1, H2, S1, S2 and S3 observations, respectively.}
\label{fig:stokes}
\end{figure*}

\subsection{Polarimetric Analysis} \label{subsec:polarimetry}
We measured the model-independent polarization of Cygnus X--1 using \textsc{pcube} algorithm. Figure~\ref{fig:stokes} shows the contour plots of the Stokes parameter, Q/I, and U/I for all five observations in the $2-8$ keV energy band in the left panel. The right panel of Figure~\ref{fig:stokes} shows the PD and PA for all five observations. We found the PD $=4.09\pm0.25\%$ and $3.88\pm0.23 \%$ in the hard state during H1 and H2 observations, respectively. The observed PD is higher than the minimum detectable polarization (MDP) at 8.4$\sigma$. The PA is found to be $-21^\circ{}.0\pm2.7$ and $-22^\circ{}.1\pm2.5$ during H1 and H2, respectively. In the soft state observations, we measured PD as $2.49 \pm 0.25 \%$, $2.62\pm0.23 \%$, and $2.79\pm 0.21 \%$ during S1, S2, and S3, respectively. The corresponding PA was found to be $-18^\circ{}.5 \pm3.3$, $-21^\circ{}.7\pm3.6$ and $-20^\circ{}.2\pm3.4$, during S1, S2 and S3, respectively. The details result is tabulated in Table~\ref{tab:pol}.

\begin{figure}
\centering
\includegraphics[width=8.5cm]{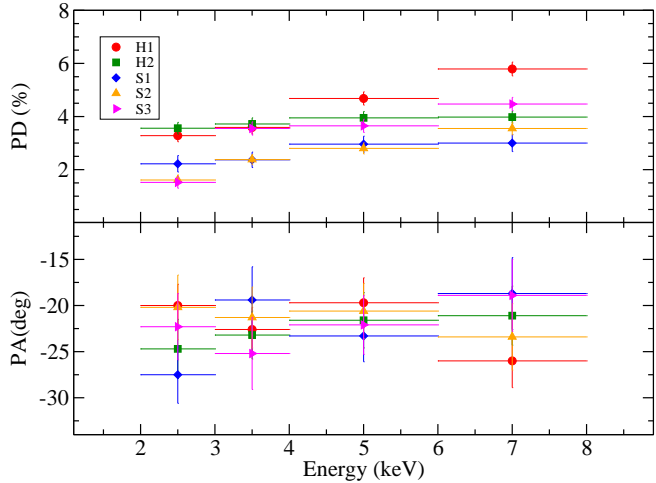}
\caption{The variation of the polarization degree (PD; \%) and polarization angle (PA; in degree) are shown as a function of energy in the upper and bottom panels. The red circles, green squares, blue diamonds, orange upper triangles, and magenta right-triangles represent the data for the H1, H2, S1, S2, and S3, respectively.}
\label{fig:en-pol}
\end{figure}

We also measured the polarization in different energy bands to check the energy-dependent polarization. We extracted the Stokes parameter in four energy bands: $2-3$ keV, $3-4$ keV, $4-6$ keV, and $6-8$ keV. We found that the PD increases with the energy in both hard and soft states. The PA did not change with the energy in both spectral states. Figure~\ref{fig:en-pol} shows the energy-dependent variation of the PD and PA in the upper and bottom panels, respectively. The energy-dependent polarization result is tabulated in Table~\ref{tab:en-pol}.

\begin{table}
\centering
\caption{Results of Polarization with \textsc{pcube} algorithm}
\begin{tabular}{cccccc}
\hline
  & Q/I & U/I & PD & PA &	MDP \\
  & (\%) & (\%) & (\%) & (degree) & (\%) \\
 \hline
H1 &	$3.04\pm	0.35$ &$ -2.74 \pm	0.36$&$	4.09\pm	0.25	$&$-21.0\pm	2.7$&	0.49 \\
H2 &	$2.78\pm	0.36$ &$ -2.71 \pm	0.31$&$	3.88\pm	0.23	$&$-22.1\pm	2.5$&	0.82 \\
S1 &	$1.99\pm	0.32$ &$ -1.51 \pm  0.38$&$	2.49\pm	0.15	$&$-18.5\pm	3.3$&	1.19 \\
S2 &	$1.91\pm	0.34$ &$ -1.82 \pm	0.31$&$	2.62\pm	0.23	$&$-21.7\pm	3.6$&	0.75 \\
S3 &	$2.12\pm	0.35$ &$ -1.81 \pm	0.23$&$	2.79\pm	0.21	$&$-20.2\pm	3.4$&	0.78 \\
\hline
\end{tabular}
\label{tab:pol}
\end{table}

\subsection{Spectral Analysis} \label{subsec:spec}
The spectral analysis is carried out in \textsc{XSPEC} v12.10. Cygnus X--1 was observed simultaneously by NICER, IXPE, and NuSTAR twice in the hard state (H1 \& H2). Although NuSTAR observed Cygnus X--1 during S3, the data was publicly unavailable at the time of writing the paper. The S1 and S2 observations were carried out only by IXPE. We started our analysis by fitting the $2-8$ keV IXPE I spectra with phenomenological models. The hard state spectra (H1 \& H2) were fitted with an absorbed powerlaw model. For absorption, we used \textsc{tbabs} model with \textsc{wilm} abundances \citep{Wilms2000}. We also required a Gaussian component for the Fe K$\alpha$ line emission. The model read is \textsc{xspec} as \textsc{tbabs*(powerlaw+gaussian)} (hereafter, Model-1). The fit returned with $\chi^2/$degrees of freedom (dof) = 1396/1332 and 1419/1332 for H1, and H2, respectively. The hydrogen column density ($N_{\rm H}$) is found to be constant across two observations, with $N_{\rm H} \sim 5.3-5.6 \times 10^{21}$ cm$^{-2}$. The photon index ($\Gamma$) was observed to increase from $\Gamma=1.53\pm0.02$ in H1 to $\Gamma=1.66\pm0.03$ in H2.

Thermal emission was detected in the IXPE energy range during the soft state. We added \textsc{diskbb} model \citep{Makishima1986} for the thermal emission. The model reads in \textsc{xspec} as \textsc{tbabs*(diskbb+powerlaw)} (hereafter Model-2). We found the inner disk temperature, $T_{\rm in}$ $\sim 0.4$~keV, with the $\Gamma$ varied in the range of $\Gamma \sim 2.5-2.9$. We also calculated the fraction of thermal emission in the soft state as $f_{\rm disk}=F_{\rm disk}/F_{\rm tot}$, where $F_{\rm disk}$ and $F_{\rm tot}$ are the disk and total flux in the $2-8$ ~keV ranges, respectively. We found $f_{\rm disk} \sim 20\%$ in all three observations.

Next, we fitted NICER+IXPE+NuSTAR spectra in a broad energy range of $1-78$~keV in the hard state. As the reprocessed emission is visible in the $10-40$~keV energy range, we replaced \textsc{powerlaw} model with \textsc{relxillcp} \citep{Garcia2014,Dauser2016} in the Model--2. We also remove the Gaussian component, as Fe K$\alpha$ line is calculated self-consistently in the \textsc{relxillcp}. The full model read in \textsc{xspec} as \textsc{const*tbabs*(diskbb+relxillcp)}. During fitting, we froze the value of the spin parameter at $0.998$, inner disk radius ($R_{\rm in}$) at 1 $R_{\rm ISCO}$. We obtained a good fit for both observations, with the fit returned as $\chi^2$/dof= 3668/3656 and 3712/3659 for H1 and H2, respectively. We obtained $T_{\rm in}\sim 0.15$~keV, $\Gamma = 1.53\pm0.04$ and $1.66\pm0.03$, $kT_{\rm e}=127^{+67}_{-23}$ and $122^{+74}_{-28}$~keV for H1 and H2, respectively.

\begin{figure*}
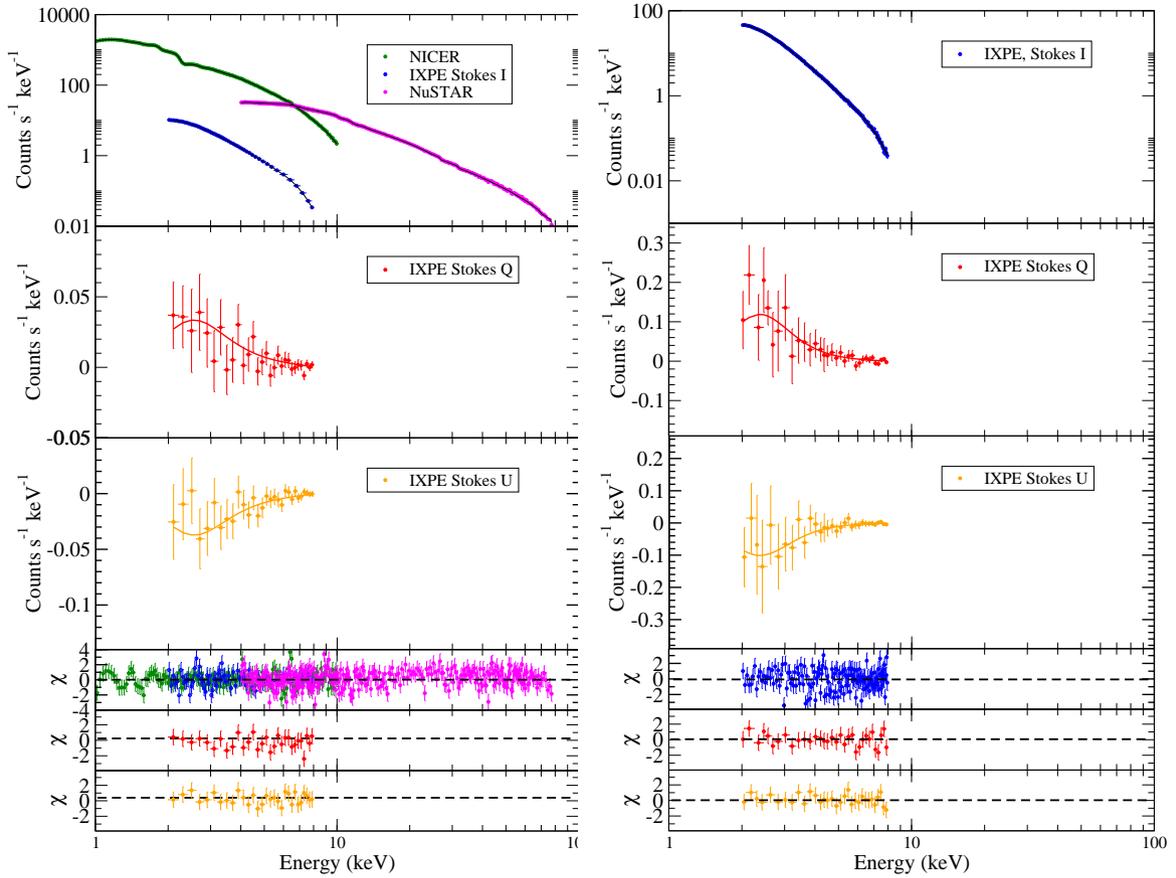

\centering
\includegraphics[width=7.7cm]{fig3a.eps}
\includegraphics[width=7.5cm]{fig3b.eps}
\caption{Left panel: Broad-band spectra in $1-78$~keV energy range shown in the top panel for H2. The green, blue and magenta points represent the NICER, IXPE (Stokes I) and NuSTAR data, respectively. In the second and third panel, the Stokes Q (red points) and Stokes U (orange points) spectra are shown. At the bottom three panels, the residuals of the best fitted model is shown. Right panel: in the top three panels, the Stokes I, Stokes Q and Stokes U spectra are shown, obtained with IXPE in S3. The bottom three panels show the residuals of the best-fit model. }
\label{fig:spec}
\end{figure*}

\subsection{Spectro-Polarimetric Analysis} \label{subsec:spec-pol}

We carried out spectro-polarimetric analysis using IXPE I, Q and U along with NICER spectra for H1 and H2. However, for S1, S2 and S3, the spectro-polarimetric analysis was carried out by simultaneously fitting the Stokes I, Q, and U spectra from the IXPE in the $2-8$~keV energy ranges, as only IXPE data was available. We used the constant polarization model \textsc{polconst} for the spectro-polarimetric analysis in \textsc{xspec}. For the hard state observation, our model reads in \textsc{xspec} as \textsc{const* tbabs * polconst (power-law + gaussian)}. The fitting result returned as PD = $3.78\pm0.45\%$ and $3.57\pm0.67$; PA = $-23^\circ{}.8\pm3.2$ and $-24^\circ{}.7\pm3.0$, for $\chi^2$/dof= 1356/1330 and 1396/1330 for H1 and H2, respectively. In the soft state, the model reads as \textsc{const* tbabs * polconst (diskbb+ power-law + gaussian)}. The PD and PA are obtained in the range of $2.2 - 2.8 \%$ and $-22^\circ{}$ to $-18^\circ{}$. 

As we found energy-dependent PD in Section~\ref{subsec:polarimetry}, we employed \textsc{polpow} and \textsc{pollin} model for the spectro-polarimetric analysis, replacing \textsc{polconst} model. The \textsc{polpow} gives PD(E)$ = PD_{\rm norm} \times E^{\rm -\alpha_{PD}}$ and PA(E)$ = \psi_{\rm norm} \times E^{-\alpha_{\rm \psi}}$. Here $PD_{\rm norm}$ is PD at 1 keV, $\alpha_{\rm {PD}}$ is index, $\psi_{\rm norm}$ is PA at 1 keV, and $\alpha_{\rm \psi}$ is PA index. As PA did not change with the energy, we fixed the index of PA, $\alpha_{\psi}$ at zero. With the \textsc{polpow} model, the fit improved statistically over the \textsc{polconst} model. The details result is quoted in Table~\ref{tab:xspec_pol}. In all observations, we obtained $\alpha_{\rm PD} \sim -0.5$ to $-0.6$. Integrating the PD over $2-8$ keV energy range, we obtained the PD as $\sim 4.1\%$, $\sim 3.8\%$, $\sim 2.4\%$, $\sim 2.4\%$ and, $\sim 2.7\%$ for H1, H2, S1, S2, and S3, respectively. These values are consistent with the PD obtained with the \textsc{pcube} algorithm.

Next, we employed \textsc{pollin} model, replacing the \textsc{polpow}. The \textsc{pollin} model describes linear dependency of PA and PD with energy. The model gives PD(E) = $A_{\rm 1}+(E-1)\times m_{\rm A}$ and PA(E) = $\psi_{\rm 1}+(E-1)\times m_{\rm \psi}$, where $A_{\rm 1}$ is PD at 1 keV, $m_{\rm A}$ slope of PD, $\psi_{\rm 1}$ PA at 1 keV, and $m_{\rm \psi}$ slope of PA. We fixed the slope of PA ($m_{\psi}$) at zero during our analysis. The fit did not improve much with \textsc{pollin} model over the \textsc{polconst} model. The detailed result of this model is tabulated in Table~\ref{tab:xspec_pol}.

\begin{table*}
\centering
\caption{Spectro-Polarimetric Results of IXPE observation using \textsc{polconst}, \textsc{polpow} and \textsc{pollin} model.}
\begin{tabular}{cccccccc}
\hline
&	&	H1			&	H2			&	S1			&	S2			&	S3			\\
 \hline
TBABS&$N_{\rm H}$	($10^{22}$ cm$^{-2}$) &$	0.56	\pm	0.02	$&$	0.53	\pm	0.02	$&$	0.59	\pm	0.03	$&$	0.50	\pm	0.04	$&$	0.52	\pm	0.03$	\\
DISKBB&$T_{\rm in} $(keV)	&	--			&	--			&$	0.41	\pm	0.02	$&$	0.40	\pm	0.02	$&$	0.43	\pm	0.02$	\\
&Diskbb Norm (10$^4$)	&	--			&	--			&$	3.78	\pm 0.43	$&$	1.91	\pm 0.31		$&$	1.61	\pm	0.28 $	\\
POWERLAW&$\Gamma$	&$	1.53	\pm	0.02	$&$	1.68	\pm	0.02	$&$	2.56	\pm	0.04	$&$	2.61	\pm	0.03	$&$	2.91	\pm	0.06$	\\
\hline
FLUX&$F_{\rm 2-8~keV}^*$ 	&$	5.91\pm	0.06	$&$	5.84\pm	0.05	$&$	10.22	\pm	0.05	$&$	14.03	\pm	0.08	$&$	14.98	\pm 0.07 $		\\
&$f_{\rm disk}$	&	--			&--			&$	0.21	\pm	0.02	$&$	0.19	\pm	0.02	$&$	0.22	\pm	0.02$	\\
\hline
POLCONST&PD (\%)	&$	3.82	\pm	0.26	$&$	3.67	\pm	0.35	$&$	2.58	\pm	0.29	$&$	2.81	\pm	0.26	$&$	2.83	\pm	0.22$	\\
&$\psi$ (deg)	&$	-23.0	\pm	2.4	$&$	-24.7	\pm	2.1	$&$	-23.5	\pm	2.9	$&$	-21.3	\pm	2.5	$&$	-20.5	\pm	3.1$	\\
&$\chi^2/$dof	&	2712/2689	&	2722/2687	&	1350/1332	&	1357/1332	&	1399/1332\\
\hline
POLPOW&PD$_{\rm norm}$ (\%)	&$	1.62	\pm	0.27	$&$	1.41	\pm	0.11	$&$	0.93	\pm	0.18	$&$	1.04	\pm	0.18	$&$	1.14	\pm	0.18$	\\
&$\alpha_{\rm PD}$ & $-0.57\pm0.06$ & $-0.53\pm0.05$ & $-0.52\pm0.06$ & $-0.57\pm0.07$ & $-0.53\pm0.05$\\
&$\psi$ (deg)	&$	-23.0	\pm	2.4	$&$	-24.7	\pm	2.1	$&$	-23.3	\pm	3.1	$&$	-22.5	\pm	2.8	$&$	-20.8	\pm	3.4$	\\
&$\chi^2/$dof	&	2601/2686			&	2635/2684			&	1292/1329	&	1286/1329	&	1302/1329	\\
\hline
POLLIN&PD$_{\rm 1}$ (\%)&$	1.99\pm	0.34$&$	1.89\pm	0.19	$&$	1.06\pm	0.13$&$	1.15\pm	0.21$&$	1.18\pm	0.24$\\
&$m_{\rm PD}$ (10$^{-3}$)& $5.11\pm0.51$ & $5.15\pm0.66$ & $5.78\pm0.81$ & $6.01\pm0.88$ & $5.75\pm0.81$ \\
&$\psi$ (deg)&$-23.0\pm2.6	$&$	-24.8\pm2.3	$&$	-23.7\pm3.5	$&$	-21.8\pm3.0	$&$ -20.6\pm2.8$	\\
&$\chi^2/$dof	&	2717/2686	&	2700/2684	&	1341/1329 &	1351/1329	&	1381/1329	\\
\hline
\end{tabular}
\label{tab:xspec_pol}
\leftline{$^*F_{\rm 2-8~keV}$ in the unit of $10^{-9}$ erg cm$^{-2}$ s$^{-1}$. Errors are quoted at 1.6$\sigma$ level.}
\end{table*}

\section{Discussion and Conclusion}\label{sec:dis}

We report the polarization properties of Cygnus X--1 in the $2-8$~keV energy ranges using the data obtained by IXPE. We studied polarization in both hard and soft spectral states. For the first time, we reported the polarization of the source in the soft spectral state.

One needs to study both spectral and timing properties to identify the spectral states. The variation of MAXI count rate and HR clearly shows that the 2022 and 2023 observations were made in different states. The spectral analysis results also supported this. The 2022 and 2023 spectra are consistent with the hard and soft states, for Cygnus X–1, respectively, as previously reported \citep{Yan2021,Walton2016}. Hence, we did not study the timing properties in detail in the present work.

We calculated the PD and PA from two different approaches, one with the \textsc{pcube} algorithm, and other by the spectro-polarimetric analysis. We obtained a consistent results from both methods within 3$\sigma$ level. In \textsc{xspec}, both constant (\textsc{polconst}) and energy-dependent polarization (\textsc{polpow} \& \textsc{pollin}) model gave us a good fitting although \textsc{pcube} suggested energy-dependent PD. This could be due to low-SNR of the IXPE Q and U spectra. However, the fitting with the \textsc{polpow} model returned with the best-fit.

Based on \textsc{monk} code, \citet{Ursini2022} predicted the PD to be $\leq 1\%$, $\sim 1-4 \%$, and $\sim 4\%$ for a low inclination system for the spherical lamp-post, conical and slab geometry, respectively. Depending on the inclination angle, the PD could be $\sim 15\%$ for a slab corona, while a conical corona can yield PD up to $\sim 7\%$. For the lamp-post geometry, the PD would be in the range of $\sim 1-3\%$. \citet{Poutanen2023} predicted the PD could be around $\sim 5\%$ for a system with inclination angle of $\sim 30^{\circ{}}$ for an outflowing corona. \citet{Dexter2023} proposed that the polarization arises from the bulk Comptonization in a mildly relativistic jet in Cygnus X--1 in the hard state.

We estimated the PD $\sim 4\%$, and $\sim 2.5-2.8\%$ in the hard and soft state, respectively, in the $2-8$~keV energy range. The PA is constant within uncertainty across five observations, with $\psi \sim 18^\circ{}-22^\circ{}$. Recently, \citet{Krawczynski2022} also studied Cygnus X--1 using the same IXPE data in the hard state and found similar PA, with the PA aligned along the radio jet axis. A similar PA in the hard and soft states indicates that the X-ray-emitting plasma is perpendicular to the radio jet axis. The observed PD suggested that the coronal geometry could be conical or slab in the hard state. However, the extended conical geometry would be preferred if the coronal geometry remains the same in both spectral states. The polarimetric result also rejected a lamp-post corona.

The PD increases with the energy in all observations; however, the PA did not change. A similar trend of increasing PD with energy is seen in other X-ray binaries and active galactic nuclei \citep{Rawat2023,Chatterjee2023}. Various numerical simulations predicted a similar trend of increasing PD with the energy in the $2-8$~keV energy range \citep{Krawczynski2012,Ursini2022}.

The joint fitting of NICER+IXPE+NuSTAR spectra in the $1-78$~keV energy range in the hard state suggested a disk of temperature $T_{\rm in}\sim 0.15$ keV, $\Gamma \sim 1.6$, and $kT_{\rm e} \sim 130$ keV. We studied the IXPE spectra in the soft state and found a hotter disk with an inner disk temperature of $\sim 0.4$~keV and the photon index $\Gamma \sim 2.6-2.9$. The spectral analysis suggested that the disk contributes $<1\%$ in the hard state and $\sim 20\%$ in the soft state, in the $2-8$~keV energy range. The iron line emission contributed $<10\%$ in the hard and soft states in the $2-8$~keV energy bands.

Recently, IXPE measures the PD as $\sim 8\%$ at the $2-8$~keV energy ranges for in BHXB 4U 1630--47 \citep{Kushwaha2023,Rawat2023} in the high soft state. As only the disk contributes to the entire energy spectrum, the polarization is attributed to the disk emission \citep{Rawat2023}. In the steep power-law state (SPL), the PD is observed to decrease to $\sim 7\%$, with the appearance of a weak corona. \citet{Rawat2023b} argued that the corona could depolarize which led to decrease in the PD in the SPL. \citet{Majumder2023} found a PD of $\sim 3\%$ in LMC X--3 in the soft state. In both sources, the observed PD is higher than Cygnus X--1 in the soft state. This is expected as both 4U 1630--47 ($i\sim75^\circ{}$) and LMC X--3 ($i\sim69^\circ{}$) are high inclination sources, contrary to Cygnus X--1, which is a low inclination source \citep{Schnittman2009}. A similar polarization properties is seen in black hole candidate Swift J1727.8--1613. The PD decreases from $\sim 4\%$ to $\sim 3\%$ as the source moves to the HIMS from the LHS \citep{Ingram2023}. The author argued that the increasing disk emission reduces the PD, as the the polarization of the disk could be perpendicular to that of the corona \citep[e.g.,][]{Ingram2023}. On the other hand, surprisingly, no polarization is detected in LMC X--1, where thermal disk emission dominate \citep{Podgorny2023}.

As the PD decreases for the coronal emissions, the scattering inside the corona is likely to reduce the polarization. This could also explain the different PD in the hard and soft states. The PD decreases in the soft state. In the hard state, the corona is hot and optically thin. Thus, seed photons suffer less scattering inside the corona before escaping. However, the corona is cooler, dense, and optically thick in the soft state. Hence, a seed photon would suffer a high number of scattering inside the corona before escaping it \citep[e.g.,][]{ST80,Chatterjee2017a,Chatterjee2017b,Jana2023}. As the corona could depolarize, it would be related to the number of scattering of the seed photons \citep[e.g.,][]{Rawat2023b}; hence, the more number of scattering in an optically dense corona would reduce the PD in the soft state. Another possibility is that the polarization of the disc is perpendicular to that of the corona. As the disc emission rises in the soft state, the PD could decrease \citep[e.g.,][]{Krawczynski2023,Ingram2023}.

In this paper, we studied the polarization properties of Cygnus X--1 in the hard and states. We find the PD decreases in the soft state, although the PA remains the same. We proposed that the PD depends on the number of scattering in the corona, and it can explain the different PD in hard and soft states. On the other hand, if the disk is polarized perpendicular to that of the corona, it could also decrease the PD in the soft state. In the future, we will study Cygnus X--1 in a broad energy range in both hard and soft states to probe the polarization properties.

\section*{Acknowledgments}

AJ and HK acknowledge the support of the grant from the Ministry of Science and Technology of Taiwan with the grand numbers MOST 110-2811-M-007-500 and MOST 111-2811-M-007-002. HK acknowledge the support of the grant from the Ministry of Science and Technology of Taiwan with the grand number MOST 110-2112-M-007-020 and MOST-111-2112-M-007-019. We alos thank the IXPE, NuSTAR and NICER team for providing the data and software packages.

\section*{Data Availability}
All the data are publicly available.



\bibliographystyle{mnras}
\bibliography{cyg_x1} 

\begin{thebibliography}{}
\makeatletter
\relax
\def\mn@urlcharsother{\let\do\@makeother \do\$\do\&\do\#\do\^\do\_\do\%\do\~}
\def\mn@doi{\begingroup\mn@urlcharsother \@ifnextchar [ {\mn@doi@}
  {\mn@doi@[]}}
\def\mn@doi@[#1]#2{\def\@tempa{#1}\ifx\@tempa\@empty \href
  {http://dx.doi.org/#2} {doi:#2}\else \href {http://dx.doi.org/#2} {#1}\fi
  \endgroup}
\def\mn@eprint#1#2{\mn@eprint@#1:#2::\@nil}
\def\mn@eprint@arXiv#1{\href {http://arxiv.org/abs/#1} {{\tt arXiv:#1}}}
\def\mn@eprint@dblp#1{\href {http://dblp.uni-trier.de/rec/bibtex/#1.xml}
  {dblp:#1}}
\def\mn@eprint@#1:#2:#3:#4\@nil{\def\@tempa {#1}\def\@tempb {#2}\def\@tempc
  {#3}\ifx \@tempc \@empty \let \@tempc \@tempb \let \@tempb \@tempa \fi \ifx
  \@tempb \@empty \def\@tempb {arXiv}\fi \@ifundefined
  {mn@eprint@\@tempb}{\@tempb:\@tempc}{\expandafter \expandafter \csname
  mn@eprint@\@tempb\endcsname \expandafter{\@tempc}}}

\bibitem[\protect\citeauthoryear{{Baldini} et~al.,}{{Baldini}
  et~al.}{2022}]{Baldini2022}
{Baldini} L.,  et~al., 2022, \mn@doi [SoftwareX] {10.1016/j.softx.2022.101194},
  \href {https://ui.adsabs.harvard.edu/abs/2022SoftX..1901194B} {19, 101194}

\bibitem[\protect\citeauthoryear{{Basak}, {Zdziarski}, {Parker}  \&
  {Islam}}{{Basak} et~al.}{2017}]{Basak2017}
{Basak} R.,  {Zdziarski} A.~A.,  {Parker} M.,   {Islam} N.,  2017, \mn@doi
  [\mnras] {10.1093/mnras/stx2283}, \href
  {https://ui.adsabs.harvard.edu/abs/2017MNRAS.472.4220B} {472, 4220}

\bibitem[\protect\citeauthoryear{{Chakrabarti} \& {Titarchuk}}{{Chakrabarti} \&
  {Titarchuk}}{1995}]{CT95}
{Chakrabarti} S.,  {Titarchuk} L.~G.,  1995, \mn@doi [\apj] {10.1086/176610},
  \href {https://ui.adsabs.harvard.edu/abs/1995ApJ...455..623C} {455, 623}

\bibitem[\protect\citeauthoryear{{Chatterjee}, {Chakrabarti}  \&
  {Ghosh}}{{Chatterjee} et~al.}{2017a}]{Chatterjee2017a}
{Chatterjee} A.,  {Chakrabarti} S.~K.,   {Ghosh} H.,  2017a, \mn@doi [\mnras]
  {10.1093/mnras/stw2975}, \href
  {https://ui.adsabs.harvard.edu/abs/2017MNRAS.465.3902C} {465, 3902}

\bibitem[\protect\citeauthoryear{{Chatterjee}, {Chakrabarti}  \&
  {Ghosh}}{{Chatterjee} et~al.}{2017b}]{Chatterjee2017b}
{Chatterjee} A.,  {Chakrabarti} S.~K.,   {Ghosh} H.,  2017b, \mn@doi [\mnras]
  {10.1093/mnras/stx1916}, \href
  {https://ui.adsabs.harvard.edu/abs/2017MNRAS.472.1842C} {472, 1842}

\bibitem[\protect\citeauthoryear{{Chatterjee}, {Agrawal}, {Jayasurya}  \&
  {Katoch}}{{Chatterjee} et~al.}{2023}]{Chatterjee2023}
{Chatterjee} R.,  {Agrawal} V.~K.,  {Jayasurya} K.~M.,   {Katoch} T.,  2023,
  \mn@doi [\mnras] {10.1093/mnrasl/slad026}, \href
  {https://ui.adsabs.harvard.edu/abs/2023MNRAS.521L..74C} {521, L74}

\bibitem[\protect\citeauthoryear{{Chattopadhyay} et~al.,}{{Chattopadhyay}
  et~al.}{2023}]{Chattopadhyay2023}
{Chattopadhyay} T.,  et~al., 2023, \mn@doi [arXiv e-prints]
  {10.48550/arXiv.2306.04057}, \href
  {https://ui.adsabs.harvard.edu/abs/2023arXiv230604057C} {p. arXiv:2306.04057}

\bibitem[\protect\citeauthoryear{{Chauvin} et~al.,}{{Chauvin}
  et~al.}{2018}]{Chauvin2018}
{Chauvin} M.,  et~al., 2018, \mn@doi [Nature Astronomy]
  {10.1038/s41550-018-0489-x}, \href
  {https://ui.adsabs.harvard.edu/abs/2018NatAs...2..652C} {2, 652}

\bibitem[\protect\citeauthoryear{{Dauser}, {Garc{\'\i}a}, {Walton}, {Eikmann},
  {Kallman}, {McClintock}  \& {Wilms}}{{Dauser} et~al.}{2016}]{Dauser2016}
{Dauser} T.,  {Garc{\'\i}a} J.,  {Walton} D.~J.,  {Eikmann} W.,  {Kallman} T.,
  {McClintock} J.,   {Wilms} J.,  2016, \mn@doi [\aap]
  {10.1051/0004-6361/201628135}, \href
  {https://ui.adsabs.harvard.edu/abs/2016A&A...590A..76D} {590, A76}

\bibitem[\protect\citeauthoryear{{Dexter} \& {Begelman}}{{Dexter} \&
  {Begelman}}{2023}]{Dexter2023}
{Dexter} J.,  {Begelman} M.~C.,  2023, \mn@doi [arXiv e-prints]
  {10.48550/arXiv.2308.01963}, \href
  {https://ui.adsabs.harvard.edu/abs/2023arXiv230801963D} {p. arXiv:2308.01963}

\bibitem[\protect\citeauthoryear{{Done}, {Gierli{\'n}ski}  \& {Kubota}}{{Done}
  et~al.}{2007}]{Done2007}
{Done} C.,  {Gierli{\'n}ski} M.,   {Kubota} A.,  2007, \mn@doi [\aapr]
  {10.1007/s00159-007-0006-1}, \href
  {https://ui.adsabs.harvard.edu/abs/2007A&ARv..15....1D} {15, 1}

\bibitem[\protect\citeauthoryear{{Garc{\'\i}a} et~al.,}{{Garc{\'\i}a}
  et~al.}{2014}]{Garcia2014}
{Garc{\'\i}a} J.,  et~al., 2014, \mn@doi [\apj] {10.1088/0004-637X/782/2/76},
  \href {https://ui.adsabs.harvard.edu/abs/2014ApJ...782...76G} {782, 76}

\bibitem[\protect\citeauthoryear{{Ghosh}, {Banerjee}  \& {Chakrabarti}}{{Ghosh}
  et~al.}{2019}]{Ghosh2019}
{Ghosh} A.,  {Banerjee} I.,   {Chakrabarti} S.~K.,  2019, \mn@doi [\mnras]
  {10.1093/mnras/stz402}, \href
  {https://ui.adsabs.harvard.edu/abs/2019MNRAS.484.5802G} {484, 5802}

\bibitem[\protect\citeauthoryear{{Gierli{\'n}ski}, {Zdziarski}, {Poutanen},
  {Coppi}, {Ebisawa}  \& {Johnson}}{{Gierli{\'n}ski}
  et~al.}{1999}]{Gierlinski1999}
{Gierli{\'n}ski} M.,  {Zdziarski} A.~A.,  {Poutanen} J.,  {Coppi} P.~S.,
  {Ebisawa} K.,   {Johnson} W.~N.,  1999, \mn@doi [\mnras]
  {10.1046/j.1365-8711.1999.02875.x}, \href
  {https://ui.adsabs.harvard.edu/abs/1999MNRAS.309..496G} {309, 496}

\bibitem[\protect\citeauthoryear{{Gou} et~al.,}{{Gou} et~al.}{2014}]{Gou2014}
{Gou} L.,  et~al., 2014, \mn@doi [\apj] {10.1088/0004-637X/790/1/29}, \href
  {https://ui.adsabs.harvard.edu/abs/2014ApJ...790...29G} {790, 29}

\bibitem[\protect\citeauthoryear{{Haardt} \& {Maraschi}}{{Haardt} \&
  {Maraschi}}{1993}]{HM93}
{Haardt} F.,  {Maraschi} L.,  1993, \mn@doi [\apj] {10.1086/173020}, \href
  {https://ui.adsabs.harvard.edu/abs/1993ApJ...413..507H} {413, 507}

\bibitem[\protect\citeauthoryear{{Harrison} et~al.,}{{Harrison}
  et~al.}{2013}]{Harrison2013}
{Harrison} F.~A.,  et~al., 2013, \mn@doi [\apj] {10.1088/0004-637X/770/2/103},
  \href {https://ui.adsabs.harvard.edu/abs/2013ApJ...770..103H} {770, 103}

\bibitem[\protect\citeauthoryear{{Ingram} et~al.,}{{Ingram}
  et~al.}{2023}]{Ingram2023}
{Ingram} A.,  et~al., 2023, \mn@doi [arXiv e-prints]
  {10.48550/arXiv.2311.05497}, \href
  {https://ui.adsabs.harvard.edu/abs/2023arXiv231105497I} {p. arXiv:2311.05497}

\bibitem[\protect\citeauthoryear{{Jana} et~al.,}{{Jana}
  et~al.}{2023}]{Jana2023}
{Jana} A.,  et~al., 2023, \mn@doi [\mnras] {10.1093/mnras/stad2140}, \href
  {https://ui.adsabs.harvard.edu/abs/2023MNRAS.524.4670J} {524, 4670}

\bibitem[\protect\citeauthoryear{{Jourdain}, {Roques}, {Chauvin}  \&
  {Clark}}{{Jourdain} et~al.}{2012}]{Jourdain2012}
{Jourdain} E.,  {Roques} J.~P.,  {Chauvin} M.,   {Clark} D.~J.,  2012, \mn@doi
  [\apj] {10.1088/0004-637X/761/1/27}, \href
  {https://ui.adsabs.harvard.edu/abs/2012ApJ...761...27J} {761, 27}

\bibitem[\protect\citeauthoryear{{Kislat}, {Clark}, {Beilicke}  \&
  {Krawczynski}}{{Kislat} et~al.}{2015}]{Kislat2015}
{Kislat} F.,  {Clark} B.,  {Beilicke} M.,   {Krawczynski} H.,  2015, \mn@doi
  [Astroparticle Physics] {10.1016/j.astropartphys.2015.02.007}, \href
  {https://ui.adsabs.harvard.edu/abs/2015APh....68...45K} {68, 45}

\bibitem[\protect\citeauthoryear{{Krawczynski}}{{Krawczynski}}{2012}]{Krawczynski2012}
{Krawczynski} H.,  2012, \mn@doi [\apj] {10.1088/0004-637X/754/2/133}, \href
  {https://ui.adsabs.harvard.edu/abs/2012ApJ...754..133K} {754, 133}

\bibitem[\protect\citeauthoryear{{Krawczynski} et~al.,}{{Krawczynski}
  et~al.}{2022}]{Krawczynski2022}
{Krawczynski} H.,  et~al., 2022, \mn@doi [Science] {10.1126/science.add5399},
  \href {https://ui.adsabs.harvard.edu/abs/2022Sci...378..650K} {378, 650}

\bibitem[\protect\citeauthoryear{{Krawczynski}, {Yuan}, {Chen}, {Rodriguez
  Cavero}, {Hu}, {Gau}, {Steiner}  \& {Dov{\v{c}}iak}}{{Krawczynski}
  et~al.}{2023}]{Krawczynski2023}
{Krawczynski} H.,  {Yuan} Y.,  {Chen} A.~Y.,  {Rodriguez Cavero} N.,  {Hu} K.,
  {Gau} E.,  {Steiner} J.~F.,   {Dov{\v{c}}iak} M.,  2023, \mn@doi [arXiv
  e-prints] {10.48550/arXiv.2307.13141}, \href
  {https://ui.adsabs.harvard.edu/abs/2023arXiv230713141K} {p. arXiv:2307.13141}

\bibitem[\protect\citeauthoryear{{Kushwaha}, {Jayasurya}, {Agrawal}  \&
  {Nandi}}{{Kushwaha} et~al.}{2023}]{Kushwaha2023}
{Kushwaha} A.,  {Jayasurya} K.~M.,  {Agrawal} V.~K.,   {Nandi} A.,  2023,
  \mn@doi [\mnras] {10.1093/mnrasl/slad070}, \href
  {https://ui.adsabs.harvard.edu/abs/2023MNRAS.524L..15K} {524, L15}

\bibitem[\protect\citeauthoryear{{Laurent}, {Rodriguez}, {Wilms}, {Cadolle
  Bel}, {Pottschmidt}  \& {Grinberg}}{{Laurent} et~al.}{2011}]{Laurent2011}
{Laurent} P.,  {Rodriguez} J.,  {Wilms} J.,  {Cadolle Bel} M.,  {Pottschmidt}
  K.,   {Grinberg} V.,  2011, \mn@doi [Science] {10.1126/science.1200848},
  \href {https://ui.adsabs.harvard.edu/abs/2011Sci...332..438L} {332, 438}

\bibitem[\protect\citeauthoryear{{Long}, {Chanan}  \& {Novick}}{{Long}
  et~al.}{1980}]{Long1980}
{Long} K.~S.,  {Chanan} G.~A.,   {Novick} R.,  1980, \mn@doi [\apj]
  {10.1086/158027}, \href
  {https://ui.adsabs.harvard.edu/abs/1980ApJ...238..710L} {238, 710}

\bibitem[\protect\citeauthoryear{{Majumder}, {Kushwaha}, {Das}  \&
  {Nandi}}{{Majumder} et~al.}{2023}]{Majumder2023}
{Majumder} S.,  {Kushwaha} A.,  {Das} S.,   {Nandi} A.,  2023, \mn@doi [arXiv
  e-prints] {10.48550/arXiv.2309.06845}, \href
  {https://ui.adsabs.harvard.edu/abs/2023arXiv230906845M} {p. arXiv:2309.06845}

\bibitem[\protect\citeauthoryear{{Makishima}, {Maejima}, {Mitsuda}, {Bradt},
  {Remillard}, {Tuohy}, {Hoshi}  \& {Nakagawa}}{{Makishima}
  et~al.}{1986}]{Makishima1986}
{Makishima} K.,  {Maejima} Y.,  {Mitsuda} K.,  {Bradt} H.~V.,  {Remillard}
  R.~A.,  {Tuohy} I.~R.,  {Hoshi} R.,   {Nakagawa} M.,  1986, \mn@doi [\apj]
  {10.1086/164534}, \href
  {https://ui.adsabs.harvard.edu/abs/1986ApJ...308..635M} {308, 635}

\bibitem[\protect\citeauthoryear{{Miller-Jones} et~al.,}{{Miller-Jones}
  et~al.}{2021}]{Miller-Jones2021}
{Miller-Jones} J. C.~A.,  et~al., 2021, \mn@doi [Science]
  {10.1126/science.abb3363}, \href
  {https://ui.adsabs.harvard.edu/abs/2021Sci...371.1046M} {371, 1046}

\bibitem[\protect\citeauthoryear{{Oda}, {Gorenstein}, {Gursky}, {Kellogg},
  {Schreier}, {Tananbaum}  \& {Giacconi}}{{Oda} et~al.}{1971}]{Oda1971}
{Oda} M.,  {Gorenstein} P.,  {Gursky} H.,  {Kellogg} E.,  {Schreier} E.,
  {Tananbaum} H.,   {Giacconi} R.,  1971, \mn@doi [\apjl] {10.1086/180726},
  \href {https://ui.adsabs.harvard.edu/abs/1971ApJ...166L...1O} {166, L1}

\bibitem[\protect\citeauthoryear{{Orosz}, {McClintock}, {Aufdenberg},
  {Remillard}, {Reid}, {Narayan}  \& {Gou}}{{Orosz} et~al.}{2011}]{Orosz2011}
{Orosz} J.~A.,  {McClintock} J.~E.,  {Aufdenberg} J.~P.,  {Remillard} R.~A.,
  {Reid} M.~J.,  {Narayan} R.,   {Gou} L.,  2011, \mn@doi [\apj]
  {10.1088/0004-637X/742/2/84}, \href
  {https://ui.adsabs.harvard.edu/abs/2011ApJ...742...84O} {742, 84}

\bibitem[\protect\citeauthoryear{{Podgorny} et~al.,}{{Podgorny}
  et~al.}{2023}]{Podgorny2023}
{Podgorny} J.,  et~al., 2023, \mn@doi [arXiv e-prints]
  {10.48550/arXiv.2303.12034}, \href
  {https://ui.adsabs.harvard.edu/abs/2023arXiv230312034P} {p. arXiv:2303.12034}

\bibitem[\protect\citeauthoryear{{Poutanen}, {Veledina}  \&
  {Beloborodov}}{{Poutanen} et~al.}{2023}]{Poutanen2023}
{Poutanen} J.,  {Veledina} A.,   {Beloborodov} A.~M.,  2023, \mn@doi [\apjl]
  {10.3847/2041-8213/acd33e}, \href
  {https://ui.adsabs.harvard.edu/abs/2023ApJ...949L..10P} {949, L10}

\bibitem[\protect\citeauthoryear{{Rahoui}, {Lee}, {Heinz}, {Hines},
  {Pottschmidt}, {Wilms}  \& {Grinberg}}{{Rahoui} et~al.}{2011}]{Rahoui2011}
{Rahoui} F.,  {Lee} J.~C.,  {Heinz} S.,  {Hines} D.~C.,  {Pottschmidt} K.,
  {Wilms} J.,   {Grinberg} V.,  2011, \mn@doi [\apj]
  {10.1088/0004-637X/736/1/63}, \href
  {https://ui.adsabs.harvard.edu/abs/2011ApJ...736...63R} {736, 63}

\bibitem[\protect\citeauthoryear{{Rawat}, {Garg}  \& {M{\'e}ndez}}{{Rawat}
  et~al.}{2023a}]{Rawat2023b}
{Rawat} D.,  {Garg} A.,   {M{\'e}ndez} M.,  2023a, \mn@doi [\mnras]
  {10.1093/mnras/stad2327}, \href
  {https://ui.adsabs.harvard.edu/abs/2023MNRAS.525..661R} {525, 661}

\bibitem[\protect\citeauthoryear{{Rawat}, {Garg}  \& {M{\'e}ndez}}{{Rawat}
  et~al.}{2023b}]{Rawat2023}
{Rawat} D.,  {Garg} A.,   {M{\'e}ndez} M.,  2023b, \mn@doi [\apjl]
  {10.3847/2041-8213/acd77b}, \href
  {https://ui.adsabs.harvard.edu/abs/2023ApJ...949L..43R} {949, L43}

\bibitem[\protect\citeauthoryear{{Russell} \& {Shahbaz}}{{Russell} \&
  {Shahbaz}}{2014}]{Russell2014}
{Russell} D.~M.,  {Shahbaz} T.,  2014, \mn@doi [\mnras]
  {10.1093/mnras/stt2330}, \href
  {https://ui.adsabs.harvard.edu/abs/2014MNRAS.438.2083R} {438, 2083}

\bibitem[\protect\citeauthoryear{{Schnittman} \& {Krolik}}{{Schnittman} \&
  {Krolik}}{2009}]{Schnittman2009}
{Schnittman} J.~D.,  {Krolik} J.~H.,  2009, \mn@doi [\apj]
  {10.1088/0004-637X/701/2/1175}, \href
  {https://ui.adsabs.harvard.edu/abs/2009ApJ...701.1175S} {701, 1175}

\bibitem[\protect\citeauthoryear{{Schnittman} \& {Krolik}}{{Schnittman} \&
  {Krolik}}{2010}]{Schnittman2010}
{Schnittman} J.~D.,  {Krolik} J.~H.,  2010, \mn@doi [\apj]
  {10.1088/0004-637X/712/2/908}, \href
  {https://ui.adsabs.harvard.edu/abs/2010ApJ...712..908S} {712, 908}

\bibitem[\protect\citeauthoryear{{Shakura} \& {Sunyaev}}{{Shakura} \&
  {Sunyaev}}{1973}]{SS73}
{Shakura} N.~I.,  {Sunyaev} R.~A.,  1973, \aap, \href
  {https://ui.adsabs.harvard.edu/abs/1973A&A....24..337S} {500, 33}

\bibitem[\protect\citeauthoryear{{Stark} \& {Connors}}{{Stark} \&
  {Connors}}{1977}]{Stark1977}
{Stark} R.~F.,  {Connors} P.~A.,  1977, \mn@doi [\nat] {10.1038/266429a0},
  \href {https://ui.adsabs.harvard.edu/abs/1977Natur.266..429S} {266, 429}

\bibitem[\protect\citeauthoryear{{Sunyaev} \& {Titarchuk}}{{Sunyaev} \&
  {Titarchuk}}{1980}]{ST80}
{Sunyaev} R.~A.,  {Titarchuk} L.~G.,  1980, \aap, \href
  {https://ui.adsabs.harvard.edu/abs/1980A&A....86..121S} {500, 167}

\bibitem[\protect\citeauthoryear{{Tananbaum}, {Gursky}, {Kellogg}, {Giacconi}
  \& {Jones}}{{Tananbaum} et~al.}{1972}]{Tananbaum1972}
{Tananbaum} H.,  {Gursky} H.,  {Kellogg} E.,  {Giacconi} R.,   {Jones} C.,
  1972, \mn@doi [\apjl] {10.1086/181042}, \href
  {https://ui.adsabs.harvard.edu/abs/1972ApJ...177L...5T} {177, L5}

\bibitem[\protect\citeauthoryear{{Titarchuk}}{{Titarchuk}}{1994}]{T94}
{Titarchuk} L.,  1994, \mn@doi [\apj] {10.1086/174760}, \href
  {https://ui.adsabs.harvard.edu/abs/1994ApJ...434..570T} {434, 570}

\bibitem[\protect\citeauthoryear{{Tomsick} et~al.,}{{Tomsick}
  et~al.}{2014}]{Tomsick2014}
{Tomsick} J.~A.,  et~al., 2014, \mn@doi [\apj] {10.1088/0004-637X/780/1/78},
  \href {https://ui.adsabs.harvard.edu/abs/2014ApJ...780...78T} {780, 78}

\bibitem[\protect\citeauthoryear{{Ursini}, {Matt}, {Bianchi}, {Marinucci},
  {Dov{\v{c}}iak}  \& {Zhang}}{{Ursini} et~al.}{2022}]{Ursini2022}
{Ursini} F.,  {Matt} G.,  {Bianchi} S.,  {Marinucci} A.,  {Dov{\v{c}}iak} M.,
  {Zhang} W.,  2022, \mn@doi [\mnras] {10.1093/mnras/stab3745}, \href
  {https://ui.adsabs.harvard.edu/abs/2022MNRAS.510.3674U} {510, 3674}

\bibitem[\protect\citeauthoryear{{Walton} et~al.,}{{Walton}
  et~al.}{2016}]{Walton2016}
{Walton} D.~J.,  et~al., 2016, \mn@doi [\apj] {10.3847/0004-637X/826/1/87},
  \href {https://ui.adsabs.harvard.edu/abs/2016ApJ...826...87W} {826, 87}

\bibitem[\protect\citeauthoryear{{Weisskopf} et~al.,}{{Weisskopf}
  et~al.}{2022}]{Weisskopf2022}
{Weisskopf} M.~C.,  et~al., 2022, \mn@doi [Journal of Astronomical Telescopes,
  Instruments, and Systems] {10.1117/1.JATIS.8.2.026002}, \href
  {https://ui.adsabs.harvard.edu/abs/2022JATIS...8b6002W} {8, 026002}

\bibitem[\protect\citeauthoryear{{Wilms}, {Allen}  \& {McCray}}{{Wilms}
  et~al.}{2000}]{Wilms2000}
{Wilms} J.,  {Allen} A.,   {McCray} R.,  2000, \mn@doi [\apj] {10.1086/317016},
  \href {https://ui.adsabs.harvard.edu/abs/2000ApJ...542..914W} {542, 914}

\bibitem[\protect\citeauthoryear{{Yan}, {Rapisarda}  \& {Yu}}{{Yan}
  et~al.}{2021}]{Yan2021}
{Yan} Z.,  {Rapisarda} S.,   {Yu} W.,  2021, \mn@doi [\apj]
  {10.3847/1538-4357/ac0f7b}, \href
  {https://ui.adsabs.harvard.edu/abs/2021ApJ...919...46Y} {919, 46}

\bibitem[\protect\citeauthoryear{{Zhao} et~al.,}{{Zhao}
  et~al.}{2021}]{Zhao2021}
{Zhao} X.,  et~al., 2021, \mn@doi [\apj] {10.3847/1538-4357/abbcd6}, \href
  {https://ui.adsabs.harvard.edu/abs/2021ApJ...908..117Z} {908, 117}

\makeatother
\end{thebibliography}




\appendix

\section{Observation Log}
\begin{table*}
\centering
\caption{Observation Log}
\begin{tabular}{ccccccc}
\hline
ID  &  Instrument & Observation ID & Start Date & End Date & Total Exposure \\
    &             &                & (YYYY-MM-DD) & (YYYY-MM-DD) & (ks) \\
\hline
H1  & IXPE        & 01002901       & 2022-05-15 & 2022-05-21 & 242 \\
    & NICER       & 5100320101--5100320101 & 2022-05-15 & 2022-05-21 & 89 \\
    & NuSTAR      & 30702017002--30702017006 & 2022-05-18 & 2022-05-21 & 40 \\
\hline
H2  & IXPE        & 01250101       & 2022-06-18 & 2022-06-20 & 86 \\
    & NICER       & 5100320108     & 2022-06-20 & 2022-06-20 & 3 \\
    & NuSTAR      & 90802013002    & 2022-06-20 & 2022-06-20 & 13 \\
\hline
S1  & IXPE        & 02008201       & 2022-05-02 & 2022-05-03 & 21 \\
\hline
S2  & IXPE        & 02008301       & 2022-05-09 & 2022-05-09 & 31 \\
\hline
S3  & IXPE        & 02008401       & 2022-05-24 & 2022-05-24 & 25 \\
\hline
\end{tabular}
\label{tab:log}
\end{table*}



\section{Energy-Dependent Polarization}


\begin{table*}
\centering
\caption{Results of Energy-dependent Polarization with \textsc{pcube}}
\begin{tabular}{cccccc}
\hline
    &  & 2--3 keV & 3--4 keV & 4--6 keV & 6--8 KeV \\
\hline
H1 &PD(\%)&$3.28  \pm0.23 $&$3.59  \pm0.25 $&$4.68  \pm0.25 $&$5.79  \pm0.26  $\\
 & PA (degree)&$-20.0 \pm2.3  $&$-22.6 \pm2.4  $&$-19.7 \pm2.7  $&$-26.0 \pm2.9   $\\
\hline
H2 & PD(\%)&$3.56  \pm0.22 $&$3.72  \pm0.23 $&$3.95  \pm0.24 $&$3.98  \pm0.23  $\\
 & PA(degree)&$-24.7 \pm3.2  $&$-23.2 \pm2.6  $&$-21.6 \pm3.0  $&$-21.1 \pm3.2   $\\
\hline
S1 & PD(\%)&$2.22  \pm0.31 $&$2.37  \pm0.29 $&$2.96  \pm0.30 $&$3.00  \pm0.32  $\\
 & PA(degree)&$-27.5 \pm3.1  $&$-19.4 \pm3.6  $&$-23.3 \pm2.8  $&$-18.7 \pm3.9   $\\
\hline
S2 & PD(\%)&$1.61  \pm0.19 $&$2.38  \pm0.19 $&$2.80  \pm0.21 $&$3.55  \pm0.24  $\\
& PA(degree)&$-20.2 \pm3.4  $&$-21.3 \pm3.3  $&$-20.6 \pm3.0  $&$-23.4 \pm3.6   $\\
\hline
S3 & PD(\%)&$1.52  \pm0.22 $&$3.55  \pm0.25 $&$3.65  \pm0.25 $&$4.47  \pm0.25  $\\
 & PA(degree)&$-22.3 \pm3.6  $&$-25.2 \pm3.9  $&$-22.1 \pm3.2  $&$-18.9 \pm3.9   $\\
\hline
\end{tabular}
\label{tab:en-pol}
\end{table*}


\bsp	
\label{lastpage}
\end{document}